# Ti$_3$C$_2$T$_x$ MXene van der Waals gate contact for GaN high electron mobility transistors

*Chuanju Wang[1], Xiangming Xu[2], Shubham Tyagi[2], Paresh C. Rout[2], Udo Schwingenschlögl[2], Biplab Sarkar[3], Vishal Khandelwal[1], Xinke Liu[4], Linfei Gao[4], Mohamed Nejib Hedhili[5], Husam N. Alshareef[2]\*, Xiaohang Li[1]\**

Gate controllability is a key factor that determines the performance of GaN high electron mobility transistors (HEMTs). However, at traditional metal-GaN interface, direct chemical interaction between metal and GaN can result in fixed charges and traps, which can significantly deteriorate the gate controllability. In this study, Ti$_3$C$_2$T$_x$ MXene films were integrated into GaN HEMTs as the gate contact, wherein van der Waals heterojunctions were formed between MXene films and GaN without direct chemical bonding. The GaN HEMTs with enhanced gate controllability exhibited an extremely low off-state current (I$_{OFF}$) of $10^{-7}$ mA/mm, a record high I$_{ON}$/I$_{OFF}$ current ratio of ~$10^{13}$ (which is six orders of magnitude higher than conventional Ni/Au contact), a high off-state drain breakdown voltage of 1085 V, and a near-ideal subthreshold swing of 61 mV/dec. This work shows the great potential of MXene films as gate electrodes in wide-bandgap semiconductor devices.

**Keywords**

[1] Advanced Semiconductor Laboratory, Thuwal 23955-6900, King Abdullah University of Science and Technology (KAUST), Saudi Arabia. [2] Physical Science, and Engineering Division (PSE), King Abdullah University of Science and Technology (KAUST), Thuwal 23955-6900, Saudi Arabia. [3] Department of Electronics & Communication Engineering, IIT Roorkee, Roorkee, Uttarakhand 247667. [4] College of Materials Science and Engineering, Shenzhen Key Laboratory of Microscale Optical Information Technology, Guangdong Research Center for Interfacial Engineering of Functional Materials, Shenzhen University, 3688 Nanhai Ave, Shenzhen 518060, PR China. [5] Core Laboratories, King Abdullah University of Science and Technology (KAUST), Thuwal 23955-6900, Saudi Arabia. *e-mail: husam.alshareef@kaust.edu.sa, xiaohang.li@kaust.edu. Chuanju Wang and Xiangming Xu contributed equally to this work.

## 1. Introduction

Depletion-mode (D-mode) GaN high electron mobility transistors (HEMTs) with Schottky gate contact have been extensively utilized in high-voltage and high-frequency applications.[1-4] For example, GaN HEMTs have entered the fields of consumer electronics, hybrid electric vehicles, photovoltaic and power distribution systems.[5] Despite several excellent attributes like high-electron mobility and excellent thermal stability, Schottky gate GaN HEMTs suffer from high gate leakage current ($I_G$) and low drain current ON/OFF ratio ($I_{ON}/I_{OFF}$).[6-11] In particular, the high $I_G$ can limit the maximum breakdown voltage and gate-voltage swing, resulting in high power consumption and early device breakdown.[6, 8, 12] In addition, a high $I_G$ and low $I_{ON}/I_{OFF}$ brings detrimental effects to the radio frequency device performance such as reduced gain, power efficiency, and increased flicker noise.[8, 13, 14] In recent years, extensive research focus has been devoted to minimizing $I_G$ by adding a suitable gate dielectric.[6, 8, 15, 16] However, addition of a gate dielectric further weakens the gate control on the channel electrostatics and demands a higher gate bias voltage. Moreover, gate oxides bring in interface traps and fixed charges that result in degradation of gate controllability and causes threshold voltage ($V_{TH}$) instability issues.[6, 8, 15, 16]

To enhance the gate controllability of Schottky gate GaN HEMTs, high-quality interfaces between metal electrodes and GaN are required. Several two-dimensional (2D) and three-dimensional (3D) materials have been evaluated as Schottky gate electrodes in GaN HEMTs, such as Cu,[2, 17] Pt,[7] Cr,[18] Pb,[17, 19] Ir,[19] Mo,[20], ITO,[21, 22] TaN,[23] TiN[3], W,[24, 25] WN or WC[18], TiSi$_2$,[11] Au,[9], graphene,[10] MoS$_2$[26], and Ni.[27-29] However, the most extensively used gate electrode, Ni/Au bilayer film, is not CMOS-compatible. Additionally, GaN HEMTs composed of the other abovementioned gate electrodes usually suffer from high $I_G$ and high $I_{OFF}$. Typically, 3D metal films are deposited onto semiconductor substrates either by e-beam evaporation or magnetron sputtering as electrical contact. However, these high-energy metallization techniques can damage the semiconductor surfaces.[30-33] Furthermore, metal diffusion into the semiconductors is inevitable during the deposition and post-gate annealing (PGA) processes.[31-35] In addition, the formation of thin complex-compound layers at the metal-semiconductor interfaces is inevitable during the high-energy metallization processes.[31, 36] Henceforth, metal diffusion induced defects and interfacial compounds induced defect states will be regarded as the defect induced gap states (DIGS). Moreover, even without interfacial defects, the interaction between the metal and the semiconductor substrates can induce gap states in the semiconductor bandgap, which are referred to as metal-induced gap states (MIGS).[34, 37, 38] The DIGS and MIGS can act as fixed charges and traps which can significantly deteriorate the gate controllability.[32, 34, 37, 39]

In recent years, 2D metals have attracted tremendous research attention as a suitable replacement to traditional 3D electrical contact. [10, 40-42] Unlike conventional bonded heterojunction, 2D materials can form vdW heterojunctions without the constraint limits of lattice matching and processing compatibility requirements.[32, 33, 41] In particular, MXenes, emerging 2D materials with promising electronic and optical properties, have shown significant promise for application in electrochemical-energy storage,[43] optoelectronics,[41] and catalysis.[44] Compared with other kinds of MXenes like $Mo_2CT_x$, $Nb_2CT_x$, $T_2CT_x$, and $V_2CT_x$, $Ti_3C_2T_x$ has received significant attention due to its high conductivity, better chemical stability, and mature synthesis technique.[45] For example, $Ti_3C_2T_x$ MXene has been used in photodetectors, where fast response speed and large responsivity have been obtained. [42, 46] In addition, $Ti_3C_2T_x$ MXene shows wavelength and polarization dependence in light absorption, which could have potential applications in perfect lenses, sensors, and light sources.[47] Recently, MXenes have been attracting significant attention in the fields of solid-state electronics and iontronics owing to their unique properties including metallic conductivity, negatively charged surface,[48] and hydrophilicity.[49] In addition, by modifying the MXene surface with different functional groups (–F, –OH, and =O), work functions in the range of 2–8 eV can be achieved, providing tunable Schottky barrier height with semiconductor substrates.[50] The metallic conductivity (15100 S/cm)[51] of MXene films combined with their high work functions, achieved by terminating the surface with more =O groups,[52] highlights them as promising gate electrodes for use in wide-bandgap semiconductor electronic devices such as GaN high-electron mobility transistors (HEMTs). More importantly, MXene films can be spray-coated onto semiconductor substrates,[41, 42] where the weak force and vdW gap between MXene films and semiconductor substrates can suppress the DIGS and MIGS which can significantly boost the gate controllability. Here, we report an atomically-flat and ultra-clean Schottky interface morphology by integrating MXene films as the gate contact on GaN HEMTs. The vdW integration yielded a record high $I_{ON}/I_{OFF}$ ratio of ~$10^{13}$, a high off-state drain breakdown voltage of 1085 V, and a near-ideal subthreshold swing of 61 mV/dec. We also demonstrate a solution-processed gate formation process that can significantly aid the wafer-scale and micro-scale resolution patterning.

## 2. Results and Discussions

### 2.1 Direct metal and vdW MXene contact with GaN substrates

To gain insights into the impacts of the traditional metal and $Ti_3C_2T_x$ MXene gate contact on the electronic properties of GaN HEMTs, we performed density-functional theory (DFT) calculations. Owning to the difficulty of simulating heterostructures with AlGaN, GaN was adopted to simplify the DFT calculations process.[53] E-beam evaporated Ni/Au is the most widely used Schottky gate electrode in D-mode GaN HEMTs, wherein Ni forms direct chemical bonds with GaN. Contrarily, MXene films can be spay-coated on the GaN substrate, wherein a vdW gap is formed between MXene and GaN, as will be discussed in detail later. Firstly, we consider an ultraclean Ni-GaN heterointerface without interfacial defects. According to Bardeen and Heine model, at the metal-semiconductor interface, the metal can induce new gap states in the semiconductor bandgap even without interfacial defects.[38] As shown in Figure 1a, the wavefunction of the metallic atoms decaying exponentially into the substrate can induce new gap states (MIGS) in the GaN bandgap of the Ni-GaN heterostructure. As the penetration depth of the metallic wavefunction into the semiconductors is ~ 0.5 nm,[54] the MIGS can be significantly suppressed through the formation of a vdW gap at the MXene-GaN heterointerface, as shown in Figure 1b. The results of the DFT calculations in Figure 1c exhibit Ga and N states at the Fermi level, attributed to MIGS due to chemical bonding. Contrarily, Figure 1d shows for the MXene-GaN heterostructure a clean bandgap, attributed to weak interaction with the MXene. The Ga and N densities of states of a GaN slab are represented in Figure 1c and d by dashed lines for comparison. The size of the bandgap of the MXene-GaN heterostructure is comparable to that of the GaN slab. Therefore, integration of the MXene with the GaN slab is predicted to preserve the semiconducting nature of GaN. As mentioned previously, the MIGS can exert a significant effect on the gate controllability of GaN HEMTs, as shown for Ni/Au and MXene gate electrodes in Figure 1e and f, respectively. The electron flow path is also shown under the gate contact regions. Under a negative gate voltage ($V_G$), two-dimensional electron gas (2DEG) in the channel is depleted (Figure S1a and b), at the same time, reverse $I_G$ occurs where electrons transport from the gate to the channel, connecting the source and drain contact in the 2DEG channel and contributing to the $I_{OFF}$. For the Ni/Au-gate GaN HEMTs, the MIGS can provide a pathway for electrons, thus increasing the defect-assisted tunneling current ($I_G$) and the $I_{OFF}$, thereby deteriorating the gate controllability.[55, 56]

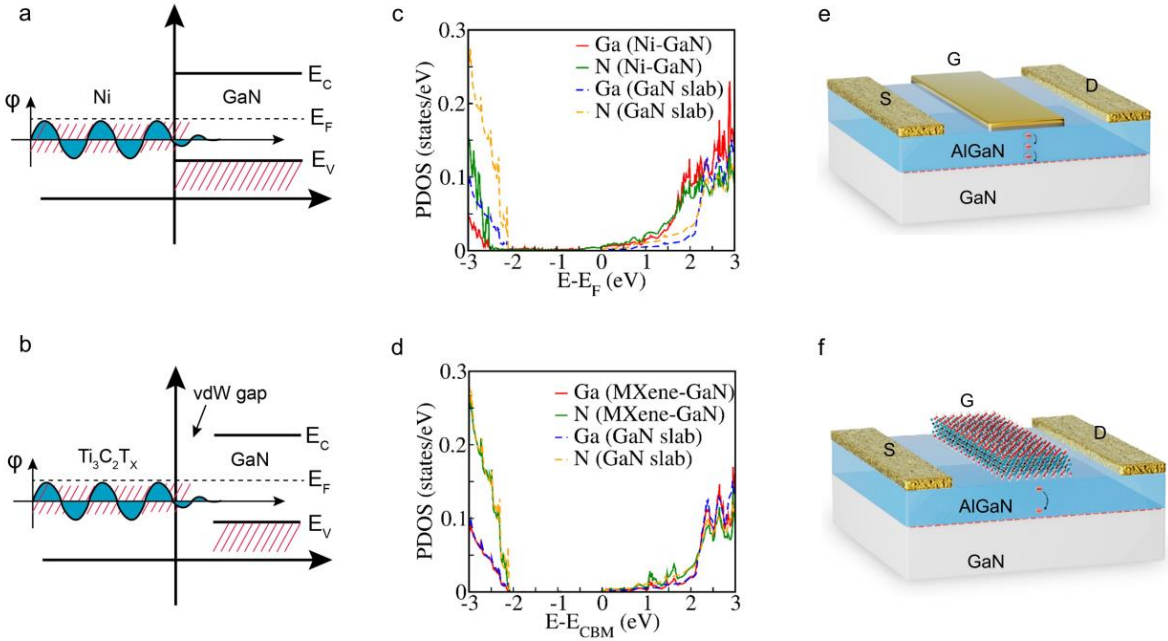

**Figure 1.** Schematic diagrams of the MIGS effect for the (a) Ni and (b) MXene contacts. (c) Ga and N densities of states (average over all Ga or N atoms) for the Ni-GaN heterostructure (solid lines) and the GaN slab (dashed lines), where the energy zero is set to the Fermi level. (d) Ga and N densities of states for the MXene-GaN heterostructure, where the energy zero is set to the conduction band minimum. The Ga and N densities of states of a GaN slab are represented by dashed lines. Schematic diagrams of the GaN HEMTs with (e) Ni/Au or (f) $Ti_3C_2T_x$ as the gate electrodes.

## 2.2 Characterization of the MXene films

$Ti_3C_2T_x$ MXene films deposition and microscale patterning processes were developed to enable MXene integration into the GaN HEMTs using standard nanofabrication techniques. Atomic force microscopy (AFM) was conducted to characterize the surface morphology of the MXene films and to measure their thickness, as depicted in the inset of Figure 2a (the thickness of a single flake was ~1.6 nm). We also prepared stacked MXene films for X-ray diffraction (XRD) characterization. The XRD patterns of the MXene films coated on GaN and glass substrates are shown in Figure 2b; the peak located at 6.6° corresponded to the (002) planes of the MXene films. The Raman spectrum in Figure 2c displays the characteristic peaks of the $Ti_3C_2T_x$ MXene at 207, 376, 575, and 739 $cm^{-1}$ which indicates the coexistence of various surface functional groups on the surface of MXene films, coinciding well with previous reports.[41] The absence of Raman peaks at 150 cm$^{-1}$ (corresponding to $TiO_2$) signifies negligible oxidization of fresh $Ti_3C_2T_x$ MXene films.[42] The XRD pattern and the Raman spectra confirm the high quality of the MXene films. The MXene films coated on GaN were annealed at 500 °C for 50 s in ultrahigh vacuum conditions ($10^{-7}$ Torr). The chemical compositions of the MXene films before and after the annealing treatment were investigated via high-resolution X-ray photoelectron spectroscopy (XPS) (Supporting Information Figure S2a–f). The XPS results revealed that the concentration of surface contaminants decreased significantly after thermal treatment (Supporting Information Figure S2b and e). Interestingly, the reduced contaminant concentration had a significant impact on the device performance, which will be discussed in the next section. The MXene colloidal suspension was then spray-coated onto the semiconductor substrates, and patterned using a lift-off process. The lift-off process was chosen because it can effectively suppress the high-energy ion bombardment of the semiconductor surface during dry etching process.[30, 32, 34] In addition, during the lift-off process, the MXene films were stable in the soft ultrasonic acetone bath, indicating their strong adhesion to the semiconductor substrate. Figure 2d shows a large area (4-inch) GaN wafer with patterned $Ti_3C_2T_x$ MXene films. Figure 2e shows the optical image of the patterned uniform MXene films. Figure 2f shows the scanning electron microscopic (SEM) image of the patterned MXene films on a 4-inch GaN wafer, with visible sharp edges. These results demonstrate that our MXene solution deposition and patterning processes are suitable for device fabrication at the wafer-scale. For radiofrequency (RF) application of GaN HEMTs, a smaller gate length (< 3 μm) is required. Besides the lift-off process, dry etching could be used to pattern the MXene films to achieve such small gate lengths.

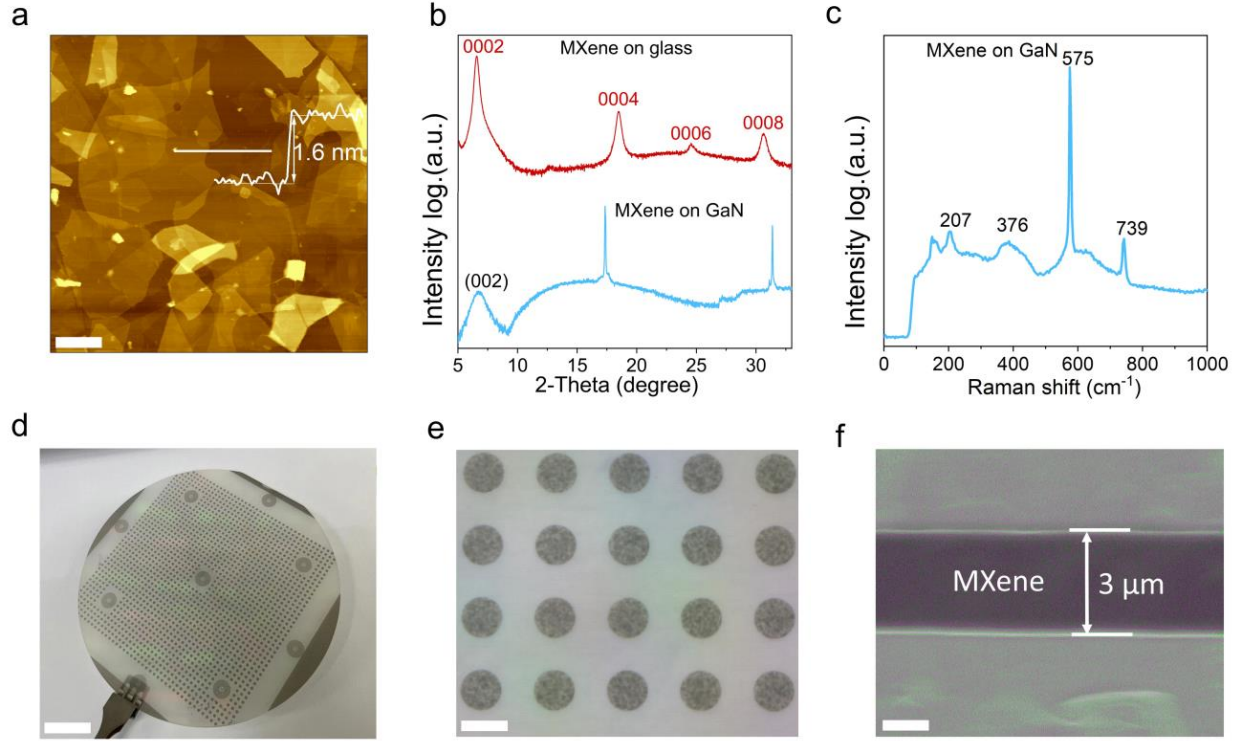

**Figure 2.** (a) AFM-height image of the $Ti_3C_2T_x$ MXene nanosheet (scale bar: 5 μm). (b) XRD pattern of the MXene films coated on glass and GaN substrates. (c) Raman spectrum of the MXene films on GaN substrate. (d) Photograph of patterned MXene films on a 4-inch GaN wafer (scale bar: 1.8 cm). (e) Optical image of the patterned MXene arrays (scale bar: 1.3 mm). (f) SEM image of the patterned MXene films (scale bar: 1.5 μm).

## 2.3. Transistors with the Ni/Au and $Ti_3C_2T_x$ MXene films as gate electrodes

Next $Ti_3C_2T_x$ MXene films were spray-coated on AlGaN/GaN epitaxial wafers and subsequently patterned as gate electrodes using the lift-off process. From the scanning transmission electron microscopy (STEM) image, the thickness of the $Ti_3C_2T_x$ MXene gate contact was measured to be ~100 nm (Figure S3a). Traditional Ni/Au-gate GaN HEMTs were also fabricated using an identical process for comparison. Figure 3a shows a schematic of the GaN HEMT with MXene as the gate contact; a vdW gap between MXene and the AlGaN layer is indicated in the figure. Figure 3b shows the SEM image of GaN HEMT arrays fabricated to explore the device uniformity and reproducibility. The SEM image was collected before electrical measurement. The length and width of the MXene gate contact are 3 and 50 μm, respectively. The SEM image of a GaN HEMT is also shown in Figure 3c, indicating the sharp edges and corners of the patterned MXene films with microscale resolution. The Ni/Au- and MXene-gate GaN HEMTs were subjected to PGA at 500 °C for 50 s under ultrahigh vacuum conditions, followed by device characterization at room temperature. A source and drain contact resistance of 0.75 Ω·mm was derived from the transmission line model (Supporting Information Figure S4). Transfer and $I_G$ curves were measured at $V_{DS}$=5V. The logarithmic-scale transfer and $I_G$ curves of the GaN HEMTs are shown in Figure 3d. It can be seen that after PGA, the $I_{ON}/I_{OFF}$ ratio was increased by one order of magnitude (Supporting Information Figure S5). Compared with the Ni/Au-gate GaN HEMTs, the MXene-gate GaN HEMTs exhibited higher $I_{ON}/I_{OFF}$ and lower maximum $I_G$ ($I_{Gmax}$) values. The linear-scale dual-sweep transfer curves of the Ni/Au- and MXene-gate GaN HEMTs are also shown in Figure 3e. Due to the vdW gap formed between the MXene films and the AlGaN layer, a superior interface contact quality is achieved. As a result, the MXene-gate GaN HEMTs exhibit a considerably smaller hysteresis during the up-sweep and down-sweep measurements. The drain current ($I_D$) versus $V_{DS}$ curves of the Ni/Au- and MXene-gate GaN HEMTs for the output current density and breakdown voltage characteristics are shown in Figure 3f-h, respectively. The two sets of GaN HEMTs exhibit no noticeable difference in the maximum $I_D$. The off-state drain breakdown voltage characteristics were measured at the gate-to-source voltage ($V_{GS}$) of −10 V, the off-state drain current of the MXene-gate GaN HEMT maintains a much lower value than the Ni/Au-gate GaN HEMT. The MXene-gate GaN HEMT also shows a larger breakdown voltage of 1085 V than that of the Ni/Au-gate GaN HEMT (770 V), signifying their great potential in high voltage applications. Further, 20 MXene-gate and 20 Ni/Au-gate GaN HEMTs were measured randomly on the wafer to explore the device uniformity and reproducibility. Figure 3i shows the dual-sweep logarithmic-scale transfer and $I_G$ curves of the GaN HEMTs after PGA. The Ni/Au-gate and MXene-gate GaN HEMTs exhibit high device-to-device uniformity. In this study, D-mode GaN HEMTs are employed, and the $V_{TH}$ of the Ni/Au-gate GaN HEMTs falls within the range of −4.5 to −5.1 V and demonstrates a narrow hump between −4.9 and −5.0 V for the 20 MXene-gate GaN HEMTs. The Schottky barrier height inhomogeneity induced by direct metal contact could be responsible for the relatively large $V_{TH}$ variations in the Ni/Au-gate GaN HEMTs.[19]

The $I_{ON}/I_{OFF}$ ratio, transconductance ($G_M$), field-effect mobility ($\mu_{FE}$), and subthreshold swing (SS) were derived from the transfer curves. Figure 3j shows the statistical-analysis data of two sets of transistors for $I_{ON}/I_{OFF}$. For the Ni/Au-gate GaN HEMTs, the $I_{Gmax}$ varies from ~$10^{-3}$ to ~$10^{-4}$ (mA/mm), and the $I_{ON}/I_{OFF}$ ratio varies from $7.5 \times 10^5$ to $8.3 \times 10^6$ for the 20 devices. The $I_{ON}/I_{OFF}$ ratios are comparable to those in previous reports obtained using Ni/Au as the contact in the D-mode Schottky gate GaN HEMTs.[4, 9, 28, 57] On the other hand, the $I_{Gmax}$ can be suppressed to lower values from ~$10^{-6}$ to ~$10^{-7}$ (mA/mm) with the MXene-gate GaN HEMTs. The $I_{ON}/I_{OFF}$ ratios ranging from $2.1 \times 10^8$ to $3.3 \times 10^9$ are two to three orders improvement compared with the Ni/Au-gate GaN HEMTs.

The statistical analysis of measured $\mu_{FE}$ was performed and shown in Figure 3k, and the calculation procedure is shown in Supporting Information Figure S6. Compared with the Ni/Au-gate GaN HEMTs, the MXene-gate GaN HEMTs exhibit ~1.5 times higher $\mu_{FE}$. The high electron mobility is effective to improve the operating frequency of the devices (Figure S6). Moreover, as shown in Figure 3l, a relatively steep SS was achieved in the MXene-gate GaN HEMTs, with a small variation from 61 to 65 mV/dec. In contrast, the Ni/Au-gate GaN HEMTs exhibit a large SS that fluctuates between 85 and 114 mV/dec. As the SS value is closely related to the gate contact quality, the steep SS of the MXene-gate GaN HEMTs indicates the lower density of states occurring at the MXene-AlGaN heterointerface. The negligible drain current-sweep hysteresis, in conjunction with the steeper SS and higher µFE, signifies the high gate controllability of the MXene-gate GaN HEMTs. The above-mentioned results verify that the MXene-gate GaN HEMTs can deliver superior performance in terms of uniformity, reproducibility, and superior transistor performance.

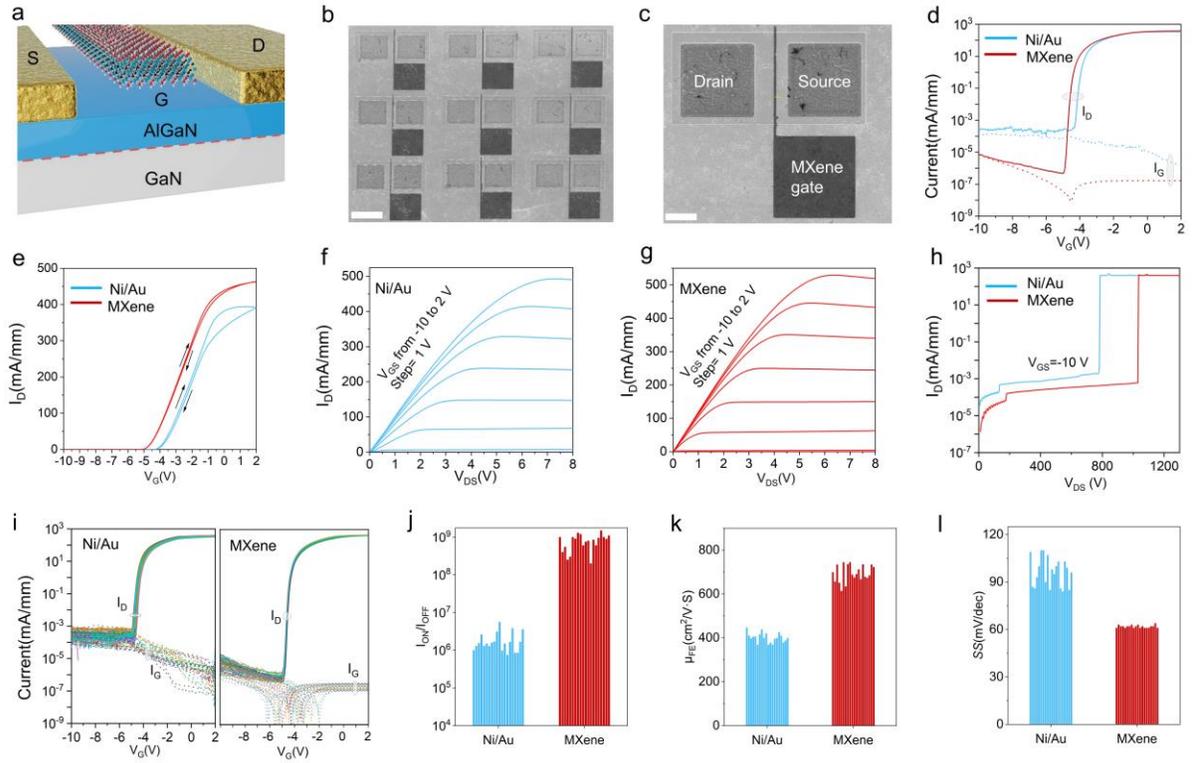

**Figure 3.** (a) Schematic diagram of a GaN HEMT with the MXene films as the gate contact. (b) SEM image of the GaN HEMT arrays (scale bar: 60 μm). (c) SEM image of a GaN HEMT (scale bar: 20 μm). (d) Logarithmic-scale transfer and $I_G$ curves of the Ni/Au- and MXene-gate GaN HEMTs. (e) Linear-scale dual-sweep transfer curves of the GaN HEMT. Output characteristics of (f) Ni/Au- and (g) MXene-gate GaN HEMTs. (h) Breakdown voltage characteristics of the GaN HEMTs. (i) Dual-sweep logarithmic-scale transfer and $I_G$ curves of the 20 GaN HEMTs. Statistical-analysis data of 20 GaN HEMTs for (j) $I_{ON}/I_{OFF}$, (k) $μ_{FE}$, and (l) SS.

**2.4 Interface characterization of Ni-AlGaN and MXene-AlGaN heterostructures**

To explore the origin of the enhanced gate controllability of the $Ti_3C_2T_x$ MXene-gate devices, STEM was conducted to characterize the interface quality of the Ni-AlGaN and MXene-AlGaN heterostructures. At the interface of the Ni-AlGaN heterostructure, amorphous-oxide layers comprising $NiO_x$ and $AlO_x$ were observed, as shown in Figure 4a.[31, 36] The nonstoichiometry-related defects located in the oxide layers can induce defect energy levels ($E_D$) in the band gap. As shown in Figure 4b, the sandwich-like MXene films display a typical 2D morphology with a thickness of ~1.6 nm for a single nanoflake. The vdW gap between the MXene films and the AlGaN layer is ~ 1.3 nm, which is close to the value reported in our previous paper.[41] In contrast with the Ni-AlGaN heterostructure, the vdW gap formed between the MXene films and the AlGaN layer retained the atomically flat and ultraclean interface morphology without interfacial oxide layers. Additionally, after vacuum annealing, the vdW gap between MXene and the AlGaN layer remained constant (Supporting Information Figure S7).

The atomic-spatial distributions of the Ni/Au-AlGaN and MXene-AlGaN heterostructures measured by electron-energy loss spectroscopy (EELS) are shown in Figure 4c and d, respectively. The results show that the high-energy Ni atoms can diffuse into the AlGaN layer during the e-beam evaporation and PGA process. [37, 58] As shown in Figure 4c, the Ni signal does not reach its minimum value before the N signal begins to increase. Therefore, intermixing of the Ni atoms and the AlGaN layer occurred at the interface. The Ni atoms that penetrated the AlGaN layer can reside at either interstitial or substitutional sites. To address their impact on the electronic structures of the GaN HEMTs, we determined the formation energies and the related defect transition levels ($q/q'$), where $q$ and $q'$ denote the charge states, by DFT calculations. The results for the Ni interstitial defect are shown in Figure 4e and the results for the Ni substitutional defects are shown in the Supporting Information Figure S8 and 9. As shown in Figure 4d, for the MXene-AlGaN/GaN heterostructure, a sharp interface was observed, signifying negligible atomic diffusion from the $Ti_3C_2T_x$ into the AlGaN layer.

At traditional metal-semiconductor interfaces, three factors can degrade the gate controllability: the MIGS, metal-induced new complex compounds at the heterointerface, and the diffusion of the metal atoms into the semiconductor substrates (the latter two factors are known as DIGS, as defined above). The MIGS and the DIGS are summarized in Figure 4e and f. The charged defects are known to act as fixed charges or traps depending on their energy levels in the bandgap.[55, 56] When the defect levels are far away from the semiconductor band edges (such as defect level (0/−1)), under either positive or negative gate bias, the Fermi level cannot move across these energy levels, they therefore behave as fixed charges. These fixed charges can scatter carriers, reducing the 2DEG mobility.[55, 56] Conversely, defect levels such as (+1/0) located close to the conduction-band minimum of GaN can induce traps, thus jeopardizing the device stability and reliability.[55, 56] Specifically, when the $V_G$ sweeps back and forth, the Fermi level is modulated to move past the defect level (+1/0), where electron trapping occurs. The electrons trapped by defect states with a large time constant do not have enough time to escape from the defect states when the $V_G$ sweeps back. Therefore, hysteresis can occur in the transfer curves, as shown in Figure 3e. The presence of the defect states can also provide a pathway for electrons to tunnel in and out of the interface region, thereby increasing the $I_G$, as shown in Figure 3d.

Capacitance-voltage (C-V) measurements with frequency ranging from 5 kHz to 1 MHz were further conducted to investigate the interface electronic properties of the Ni/Au-AlGaN/GaN and MXene-AlGaN/GaN heterostructures. In the depletion region of the capacitors, electrons are transported from the Ni/Au and MXene gate electrodes to the interfaces of Ni/Au-AlGaN/GaN and MXene-AlGaN/GaN heterostructures. The interface traps caused by the MIGS and DIGS can act as charge-trapping and detrapping centers. During the C-V measurements, the dynamic charge/discharge processes can result in capacitance variations (noise), as shown in the C-V curves in Figure 4g. In Figure 4h, the negligible noise intensity indicates the superior interface quality of the MXene-AlGaN/GaN heterostructures.

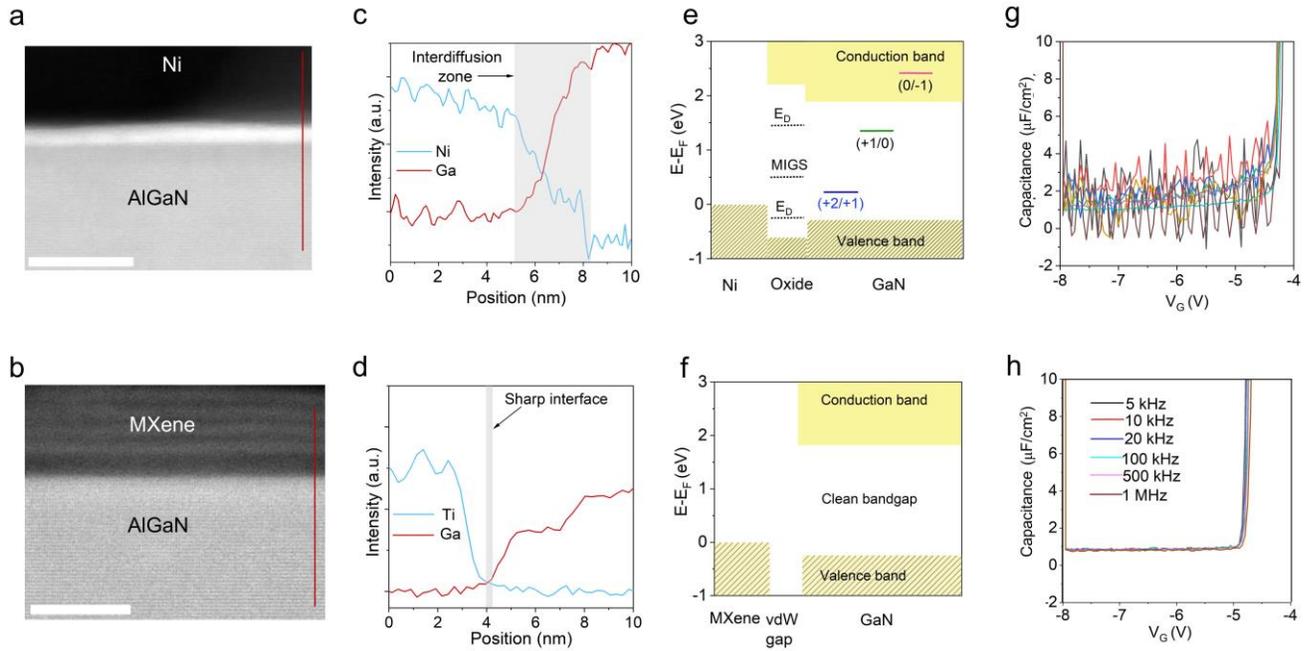

**Figure 4.** Cross-sectional STEM images of the (a) Ni-AlGaN and (b) MXene-AlGaN heterostructures (scale bar: 5 nm). EELS profiles across the interfaces from the AlGaN layer to the (c) Ni and (d) MXene films. Defect energy levels induced by the MIGS and DIGS for the (e) Ni-AlGaN and (f) MXene-AlGaN heterostructures. Frequency-dependent C-V characteristics of the (g) Ni/Au- and (h) MXene-gate GaN capacitors in the capacitor-depletion region.

## 2.5 High performance GaN HEMTs through MXene engineering

The Ti$_3$C$_2$T$_x$ MXene films were further engineered to enhance the gate controllability of GaN HEMTs. Monolayer MXene (ML-MXene) and few-layer MXene (FL-MXene) films were both tested as gate electrodes in GaN HEMTs. The AFM images of the individual flake thickness distribution for the two kinds of MXene films are shown in Supporting Information Figure S10. For the ML-MXene films, most of flake thickness was less than 2 nm, while for the FL-MXene films, individual flake thickness range over a large distribution from 5 to 20 nm. As shown in Figure 5a and b, the I$_{ON}$/I$_{OFF}$ ratio of the GaN HEMTs are $3.1 \times 10^8$ and $2.3 \times 10^{10}$ for the ML-MXene and FL-MXene, respectively. We believe that this improvement in the I$_{ON}$/I$_{OFF}$ ratio is due to the fact that MXene films with thin flakes (ML-MXene) can stack with a smaller gap between the MXene and the AlGaN layer, enabling adequate interaction between the MXene films and the AlGaN surface. However, the MXene films formed from thicker flakes (FL-MXene) would exhibit larger gaps between the MXene films and the AlGaN surface, resulting in less effective contact between MXene films and the AlGaN layer. This difference is schematically illustrated in Figure 5d and e.

In this work, ML-MXene films were further oxidized and evaluated as gate electrodes in the GaN HEMTs. The reason for this is that MXene films terminating with different surface groups have been predicted to have work functions ranging from 2 to 8 eV;[50] consequently, they can achieve a tunable barrier height with the GaN substrate. The increased content of =O surface groups would increase the work function of the MXene films, as previously predicted.[59] XPS analysis confirmed the increased surface functional =O groups after the oxidization process (Supporting Information Figure S11). As a result, a higher Schottky barrier height was obtained, as verified by the ultraviolet photoelectron spectroscopy (UPS) measurement (Supporting Information Figure S12 and S13). As shown in Figure 5f and g, the increased work function of the oxidized MXene (O-MXene) leads to a higher Schottky barrier height $\varphi_B$, which is effective for reducing I$_G$ and I$_{OFF}$. Compared with ML- and FL-MXene, the higher Schottky barrier height of the O-MXene gate contact was also confirmed by I$_G$-V$_G$ characteristics (Supporting Information Figure S14). A reasonably high barrier height observed in the O-MXene gate contact alleviates the requirement of gate dielectric which is known to degrade the gate electrostatics in channel and introduce interface defects. As shown in Figure 5c, the I$_{OFF}$ of GaN HEMTs can be significantly reduced for the O-MXene contact, and accordingly, the I$_{ON}$/I$_{OFF}$ ratio can be increased by four orders of magnitude to ~$10^{13}$, which is six orders higher than that for the traditional Ni/Au-gate GaN HEMTs.[28, 57] In addition, the ML-, FL-, and O-MXene-gate GaN HEMTS exhibit no noticeable difference in the maximum $I_D$ based on the output current density curves (Supporting Information Figure S15). Figure 5h compares the performance of GaN HEMTs with MXenes and other materials as gate electrodes.[3, 9, 10, 12, 28] The GaN HEMTs with the O-MXene gate electrodes exhibit a record high $I_{ON}/I_{OFF}$ ratio of ~$10^{13}$. To the best of our knowledge, the near-ideal SS reported in MXene gate GaN HEMTs also outperform previous reports on SS for GaN HEMTs grown on foreign substrates. However, for the transfer characteristics, we also notice that a large hysteresis occurs in the O-MXene-gate GaN HEMT during the up-sweep and down-sweep measurement (Supporting Information Figure S16a). The origin of this phenomenon is unclear, however, the combination of large hysteresis and steep SS in the case of the device using O-MXene contact is similar to devices with negative capacitance.

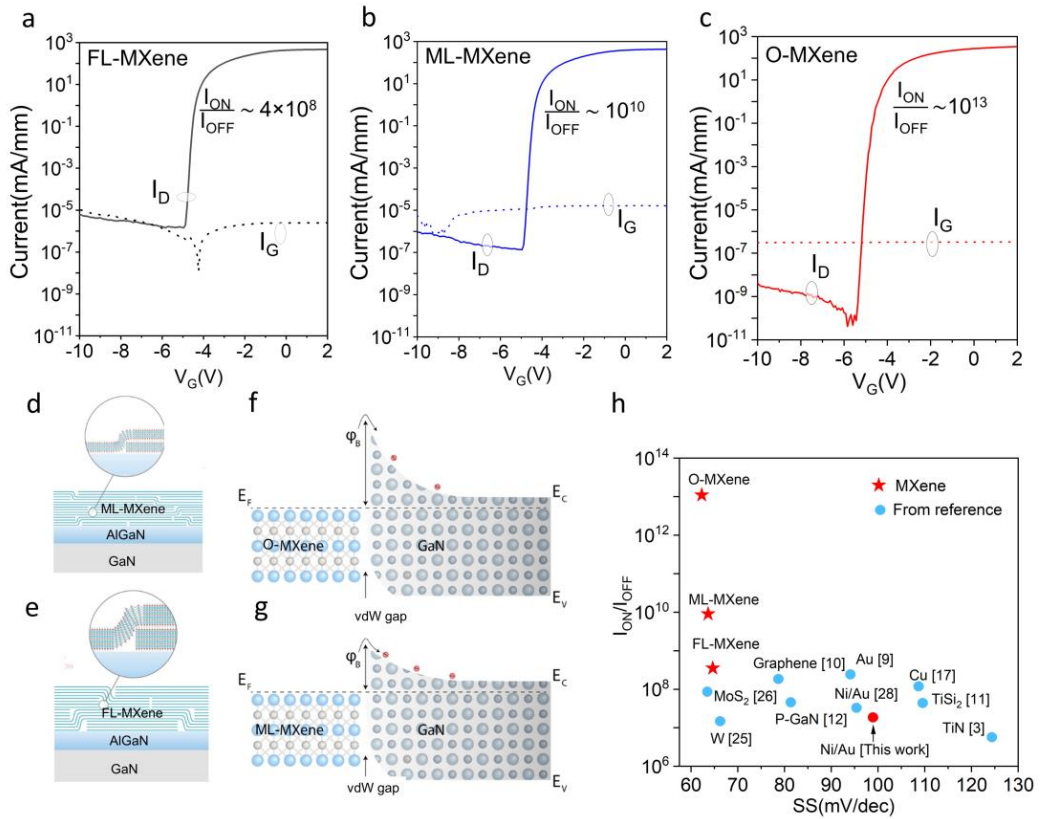

**Figure 5.** Logarithmic-scale transfer and $I_G$ curves of MXene-gate GaN HEMTs for the (a) FL-MXene, (b) ML-MXene, and (c) O-MXene gate contacts. Schematics of the (d) ML-MXene and FL-MXene (e) layer stacking. Schematics of the Schottky barrier height for the (f) O-MXene and

(g) ML-MXene films, electron flow path with a negative $V_G$ is also shown. Comparison of the $I_{ON}/I_{OFF}$ ratios and SS values of this study with reported results.

## 3. Conclusion

In summary, MXene films were employed as gate electrodes in GaN HEMTs. The vdW integration of MXene electrodes with the AlGaN/GaN epitaxial wafer can effectively avoid the MIGS and DIGS induced by traditional high-energy gate contact deposition processes. Compared with that of the traditional Ni/Au contact, the gate controllability of the MXene electrode is significantly enhanced. The MXene-gate GaN HEMTs deliver a record high $I_{ON}/I_{OFF}$ current ratio of ~$10^{13}$, high field-effect mobility, and near-ideal SS. Thus, a large breakdown voltage appended with a high $I_{ON}/I_{OFF}$ ratio and steep SS validates the high power performance of the MXene gate HEMTs. That is, the MXene gate HEMT demonstrated in this work is thought to be suitable for high power switch having a low conduction and switching loss. This work provides an Au-free, cost-effective, and low-energy integration strategy for creating gate contact not only in GaN HEMTs, but also in other GaN-based electronic and optoelectronic devices.

## 4. Experimental Section

**Structure of the AlGaN/GaN epitaxial wafer**: The AlGaN/GaN epitaxial wafer used in this study was grown on a 6-inch Si substrate by the metal-organic, chemical-vapor deposition system. A 4.7 μm GaN buffer layer was sandwiched between the Si substrate and the 300-nm GaN channel. Thereafter, a 21-nm $Al_{0.23}GaN$ barrier was deposited on top of the GaN channel.

**MXene synthesis:** The $Ti_3C_2T_x$ colloidal solution was synthesized following a previously reported method [41], after which it was stored in a vacuum ready for use.

**MXene characterizations:** XPS studies were conducted in a Kratos Axis Supra DLD spectrometer equipped with a monochromatic Al Kα X-ray source (hv = 1,486.6 eV) operating at 75 W, a multichannel plate, and delay-line detector under an ultrahigh vacuum of 1~$10^{-9}$ mbar. The survey and high-resolution spectra were collected at fixed-analyzer-pass energies of 160 and 20 eV, respectively, and quantified using empirically derived relative sensitivity factors provided by Kratos analytical. The samples were mounted in the floating mode to avoid differential charging. Charge neutralization was required for all samples. The UPS spectra were obtained using a He-I excitation (21.22 eV) source at a pass energy of 10 eV. The samples were mounted in contact mode for the UPS measurements. Raman spectra were collected using the Wintec Apyron Raman spectrometer equipment with a laser-excitation wavelength of 633 nm. The XRD patterns of the MXene films were measured using Bruker D2 PHASER. The STEM lamellae samples were fabricated by a FIB technique (Helios G4, FEI). The STEM images were acquired using the Titan G2 60-300 (FEI) instrument at an acceleration voltage of 300 kV. For the EELS measurement, the current was kept at ~41 pA to avoid beam-induced damage.

**Device fabrication and measurements:** Starting from the Mesa etching to cut off the 2DEG channel, a piranha cleaning process was deployed to fully clean the residue photoresist and remove organic contamination. Before the Ti/Al/Ti/Au deposition in the source and drain region, diluted hydrochloric acid was applied to remove the native oxide on the surface of the AlGaN/GaN epitaxial wafer. A rapid thermal annealing process at 870 °C for 50 s in Ar atmosphere was conducted to form the Ohmic contact. AZnLOF 2020 was used as the negative photoresist to define the gate region. Subsequently, the $Ti_3C_2T_x$ MXene films were spray-coated onto the AlGaN/GaN epitaxial wafer, following which the lift-off process in an ultrasonic acetone bath with controlled power was conducted to peel off the MXene films outside the gate region. The samples were subjected to ultrahigh vacuum conditions and annealed at 500 °C for 50 s. For the O-MXene gate contact, the MXene suspension was oxidized at an environment temperature of ~ 4 °C for 3 weeks. PGA was applied at 500 °C for 50 s in the $O_2$ (5 sccm)/Ar (500 sccm) mixture after the MXene patterning process. For the traditional Ni/Au-gate GaN HEMTs, after Ohmic-contact formation, PGA was conducted at 500 °C for 50 s in Ar (500 sccm). The electrical measurements were conducted using a Keysight B1500 Semiconductor Parameter Analyzer at room temperature. For the breakdown measurement, a Keysight B1505 Semiconductor Parameter Analyzer was used.

**DFT calculations:** We perform spin-polarized first-principles calculations within the framework of DFT, as implemented in the Vienna ab-initio simulation package.[60] The total energy convergence threshold is set to $10^{-6}$ eV and the structures are optimized until the Hellmann-Feynman forces stay below 0.01 eV/Å. All the calculations are performed within the Perdew-Burke-Ernzerhof generalized gradient approximation. An energy cutoff of 600 eV is used in the plane wave expansion. Electronic correlation effects in the Ni 3d orbitals are taken into account by means of an on-site Coulomb interaction of 4 eV. The Grimme method is adopted to account for the van der Waals interaction.[61] We use a $4 \times 4 \times 3$ supercell of the primitive hexagonal unit cell of GaN (dimension: 12.9 Å × 12.9 Å × 15.7 Å) to study defects. The $4 \times 4 \times 3$ supercell was used to achieve a distance larger than 10 Å between the defects in adjacent supercells (periodic boundary conditions) to avoid artificial interaction. A $1 \times 2 \times 3$ supercell of the non-primitive orthorhombic unit cell of GaN (dimension: 3.2 Å × 11.1 Å × 15.7 Å) is used to form a heterostructure with a $1 \times 3 \times 3$ supercell of Ni (dimension: 3.4 Å × 10.5 Å × 10.5 Å) in the c-direction and a $1 \times 1 \times 3$ supercell of the primitive hexagonal unit cell of GaN (dimension: 3.2 Å × 3.2 Å × 15.7 Å) is used to form a heterostructure with the unit cell of the 2D MXene $Ti_3C_2T_x$ (dimension: 3.0 Å × 3.0 Å) in the c-direction. The Ga-terminated (N-terminated) surface of each GaN slab is passivated with $H^{-0.25}$ ($H^{+0.25}$) to remove the dangling bonds and achieve a semiconducting state.[62] For comparison, we perform a separate calculation for the H-terminated $1 \times 1 \times 3$ GaN slab. The simulation cells of the heterostructures are complemented with vacuum layers of 15 Å thickness. The simulation cell of the $1 \times 1 \times 3$ GaN slab is chosen to have the same size as that of the MXene-GaN heterostructure. The calculations of defective GaN, the Ni-GaN heterostructure, the MXene-GaN heterostructure, and the H-terminated $1 \times 1 \times 3$ GaN slab are performed on Monkhorst-Pack $3 \times 3 \times 2$, $12 \times 3 \times 1$, $12 \times 12 \times 1$, and $12 \times 12 \times 1$ k-meshes, respectively. Densities of states are calculated by the tetrahedron method on Monkhorst-Pack $18 \times 5 \times 1$, $18 \times 18 \times 1$, and $18 \times 18 \times 1$ k-meshes for the Ni-GaN heterostructure, MXene-GaN heterostructure, and $1 \times 1 \times 3$ GaN slab, respectively.

**Supporting Information**

Supporting Information is available from the author.

**Acknowledgements**

Chuanju Wang and Xiangming Xu contributed equally to this work. Research reported in this publication was funded by King Abdullah University of Science and Technology.

**Conflict of Interest**

The authors declare no conflict of interest.

**Data availability**

The data that support the findings of this study are available from the corresponding author upon reasonable request.